\documentclass{nature}
\usepackage{graphicx}                           
\usepackage{times}
\usepackage{float}
\usepackage{rotating}
\usepackage{epstopdf}
\usepackage{multirow}
\usepackage{pdflscape}
\usepackage{color}

\usepackage[]{lineno}
\def\ltsima{$\; \buildrel < \over \sim \;$}
\def\simlt{\lower.5ex\hbox{\ltsima}}
\def\gtsima{$\; \buildrel > \over \sim \;$}
\def\simgt{\lower.5ex\hbox{\gtsima}}
\def\gsimeq
{\hbox{\raise0.5ex\hbox{$>\lower1.06ex\hbox{$\kern-1.07em{\sim}$}$}}}
\def\lsimeq
{\hbox{\raise0.5ex\hbox{$<\lower1.06ex\hbox{$\kern-1.07em{\sim}$}$}}}

\def\xmm{{\it XMM-Newton }}
\def\rosat{{\it ROSAT}}

\def\xmm{{\it XMM-Newton}}
\def\chandra{{\it Chandra}}

\def\fermi{{\it Fermi}}

\def\apj{ApJ}
\def\mnras{MNRAS}
\def\aap{A\&A}
\def\apjl{ApJ}
\def\apjs{ApJS}
\def\araa{ARA\&A}
\def\pasj{PASJ}
\def\nat{Nature}

\def\procspie{Proc. SPIE}
\def\gca{GeCoA}

\def\sgras{Sgr~A$^{\star}$}

\def\xis{XIS}
\def\xis1{XIS1}
\def\xis2{XIS2}
\def\xis3{XIS3}

\title{The Galactic Centre Chimney}
\author{G.~Ponti$^{1,2}$, F.~Hofmann$^{1}$, E.~Churazov$^{3,4}$, M.~R.~Morris$^{5}$, F.~Haberl$^{1}$, K.~Nandra$^{1}$, R.~Terrier$^{6}$, M.~Clavel$^{7}$, and A.~Goldwurm$^{6,8}$}

\begin{document}
\maketitle

\begin{affiliations}
 \item Max-Planck-Institut f\"ur Extraterrestrische Physik, Giessenbachstrasse 1, D-85748 Garching, Germany
 \item INAF-Osservatorio Astronomico di Brera, Via E. Bianchi 46, I-23807 Merate (LC), Italy
 \item Max-Planck-Institut f\"ur Astrophysik, Karl-Schwarzschild-Str. 1, D-85748, Garching, Germany 
 \item Space Research Institute (IKI), Profsoyuznaya 84/32, Moscow 117997, Russia
 \item Department of Physics and Astronomy, University of California, Los Angeles, CA 90095-1547, USA
 \item Unit\'e mixte de recherche Astroparticule et Cosmologie, 10 rue Alice Domon et L\'eonie Duquet, F-75205 Paris, France
 \item Univ. Grenoble Alpes, CNRS, IPAG, F-38000 Grenoble, France
 \item Service d'Astrophysique (SAp), IRFU/DSM/CEA-Saclay, F-91191 Gif-sur-Yvette Cedex, France
\end{affiliations}

\begin{abstract}
Evidence has increasingly mounted in recent decades that outflows of matter 
and energy from the central parsecs of our Galaxy have shaped the observed 
structure of the Milky Way on a variety of larger scales$^1$.   
On scales of $\sim15$~pc, the Galactic centre (GC) has bipolar lobes that 
can be seen in both X-rays and radio$^{2,3}$, indicating broadly 
collimated outflows from the centre, directed perpendicular to the Galactic plane. 
On far larger scales approaching the size of the Galaxy itself, gamma-ray 
observations have identified the so-called "Fermi Bubble" (FB) features$^4$,
implying that our Galactic centre has, or has recently had, a period of active 
energy release leading to a production of relativistic particles that now populate 
huge cavities on both sides of the Galactic plane. 
The X-ray maps from the \rosat\ all-sky survey show that the edges of these cavities 
close to the Galactic plane are bright in X-rays. 
At intermediate scales ($\sim150$~pc), radio astronomers have found the Galactic 
Centre Lobe (GCL), an apparent bubble of emission seen only at positive 
Galactic latitudes$^{5,6}$, but again indicative of energy injection from near the Galactic centre. 
Here we report the discovery of prominent X-ray structures on these intermediate 
(hundred-parsec) scales above and below the plane, which appear to connect 
the Galactic centre region to the \fermi\ bubbles. 
We propose that these newly-discovered structures, which we term the Galactic 
Centre Chimneys, constitute a channel through which energy and mass, injected 
by a quasi-continuous train of episodic events at the Galactic centre, are transported 
from the central parsecs to the base of the \fermi\ bubbles$^5$.
\end{abstract}

\begin{figure}
\begin{center}
\vspace{-2cm}
\includegraphics[width=0.7\textwidth,angle=0]{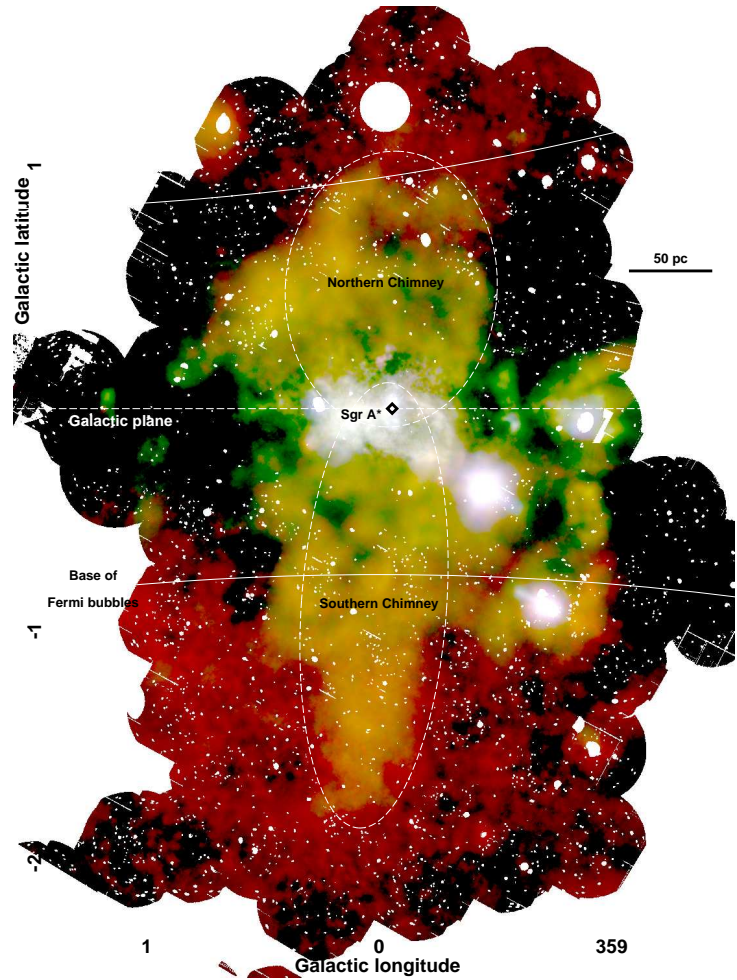}
\end{center}
\vspace{-1.7cm}
\caption{\footnotesize X-ray emission from the central $300$ by $500$ parsecs of the Milky Way.
The RGB image shows, in red: integrated emission in the energy range 1.5-2.6 keV;
green: integrated emission from 2.35 to 2.56 keV, corresponding to the S {\sc xv} transition; blue: continuum emission in the 2.7-2.97 keV band, therefore not contaminated by the intense S {\sc xv} and Ar {\sc xvii} lines $^7$. 
\sgras, the electromagnetic counterpart of the 
supermassive black hole, and the central cluster of massive stars are 
located at the centre. 
A coherent edge-brightened shell-like feature with a diameter of $\sim160$~pc, 
dubbed the "Chimney", is present north of \sgras. 
The shell is roughly co-spatial with the radio feature known as the Galactic centre lobe$^6$. Diametrically opposite to the Chimney with respect to \sgras\ 
is the newly-discovered, bright elongated feature, the southern Chimney. 
The north and south features have comparable brightness, 
extent and colour, suggesting a similar emission process (see white dashed regions). 
Their relative placement suggests a common origin located close to \sgras.
The white arcs correspond to the low-latitude edges of 
the \fermi\ bubbles, as traced by \rosat\ X-ray emission at higher latitude 
(see Fig. \ref{FourPanels}). Point sources have been removed. Larger circles correspond to regions that have been excised to remove the dust scattering haloes around bright sources. 
}
\label{fig:RGB}
\end{figure}

Figure \ref{fig:RGB} shows an X-ray map of the central $\sim300\times500$~parsec 
of the Milky Way, providing us with an unprecedented view of the Galactic corona, 
which we obtained via observations with the \xmm\ satellite between 2016 and 2018, 
in addition to archival data$^7$. 
Strong, low-energy photo-absorption that hinders the mapping of the large-scale 
diffuse emission in the GC region with \rosat\ is alleviated in this image by using 
the wider energy band of \xmm\ (see Methods). 
The Galactic supermassive black hole, \sgras, and its surrounding central cluster 
of massive stars$^8$, are located at the centre, as are the $\pm15$~pc bipolar lobes, 
visible as a barely resolved small white region perpendicular to the plane. 
The striking new features in this image are the elongated and slightly edge-brightened 
structures ($\gsimeq 1^\circ$ or $160$~pc across) to the North and South of \sgras.
We refer to these quasi linear features, which appear to connect the GC to the 
Galactic bulge, as the northern and southern Galactic Centre Chimneys. 
The northern Chimney is roughly co-spatial with the GCL. Both Chimneys share 
comparable X-ray brightness and colour. This suggests that they have a common 
origin that is most likely connected with the Galactic centre, and a common 
emission mechanism. Despite their similarities, the northern and southern Chimneys 
are not strictly symmetric about the Galactic plane, which can plausibly be attributed 
to ``Galactic weather'', i.e., the differences in the rich interstellar medium (ISM) 
structures above and below the plane that interact/interfere with the GC outflows 
of mass and energy that have presumably hewed the Chimneys.

Parallel to the plane, the breadth of the newly discovered features is limited 
to $\pm 50~$pc ($\pm 0.4^\circ$), while in the latitudinal direction, the Chimneys 
span more than a degree above and below the plane, where they merge 
with much more extended warm plasma emission seen by \rosat\ and 
tentatively identified as the X-ray counterparts of the \fermi\ bubbles 
(see Fig. \ref{FourPanels}; $^{4,9}$). 
Towards the plane, the Chimneys can be traced down to $b\sim\pm0.2^\circ$. 
Morphologically, they appear to originate from a region within $\sim50$~pc 
of \sgras, with the $\pm15$~pc bipolar lobes nested at the centre. 

\begin{figure}
\centering
\includegraphics[width=0.8\textwidth,angle=0,trim=5cm 2cm 5cm 0cm,clip=true]{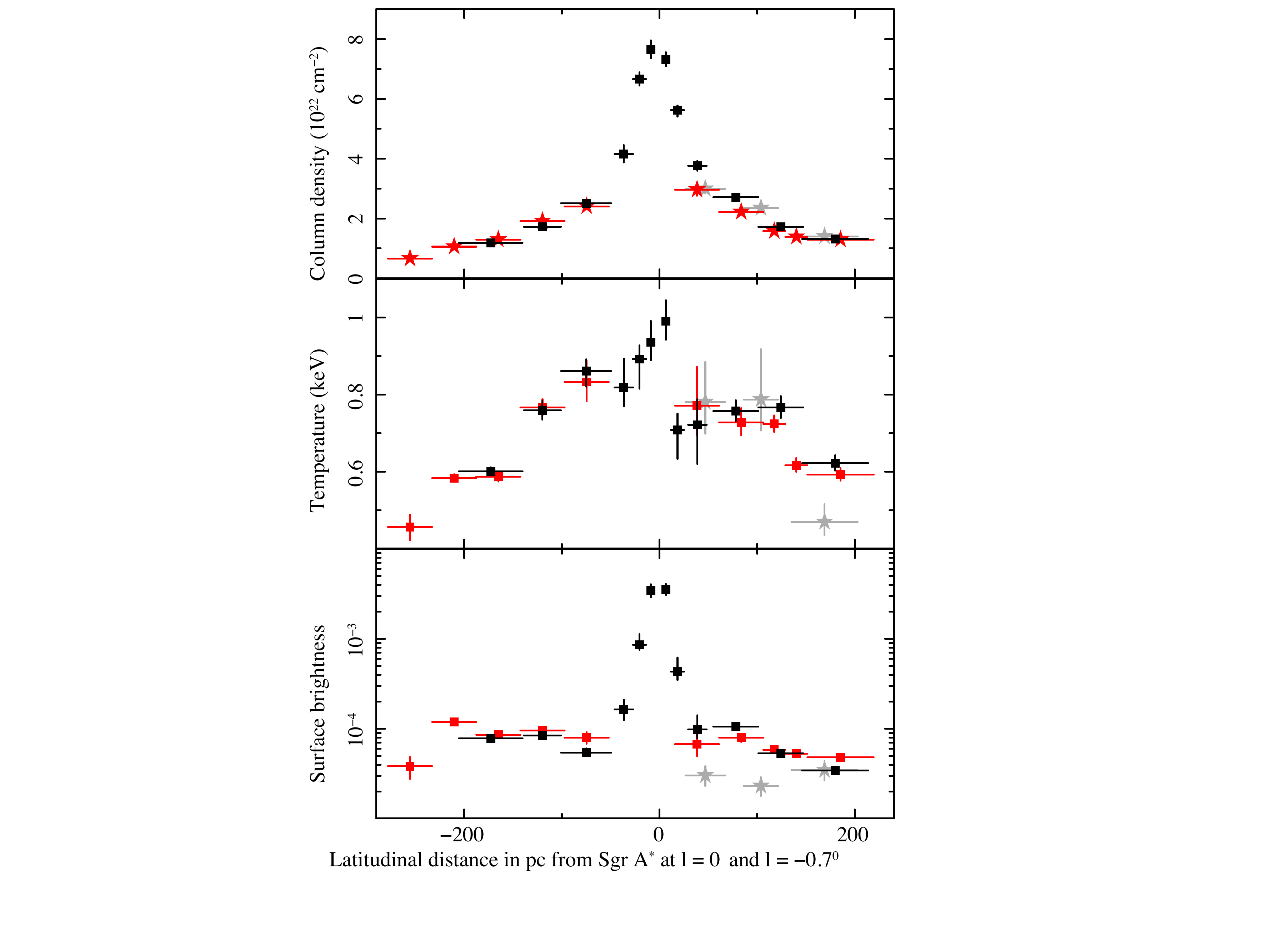}
\caption{\footnotesize Latitudinal variation of the physical parameters of the Galactic 
centre corona. Positive values indicate Galactic North. The latitude is given in parsecs, 
assuming that the emission is located at the Milky Way centre. 
From top to bottom latitudinal profiles of column density of neutral 
absorbing material, temperature and surface 
brightness (see Methods) of the X-ray emitting thermal plasma at $l=0^\circ$, 
as derived from the fit of the \chandra\ (black squares) and \xmm\ 
(red squares) spectra (see Fig. \ref{RegCha}). 
Grey stars show the latitudinal profile at $l=-0.7^\circ$ as derived 
from the fit to \xmm\ spectra (see Methods\ref{Analysis}). } 
\label{Prof}
\end{figure}

\begin{figure}
\centering
\vspace{-2cm}
\includegraphics[width=0.8\textwidth,angle=0,trim=5cm 0cm 5cm 0cm,clip=true]{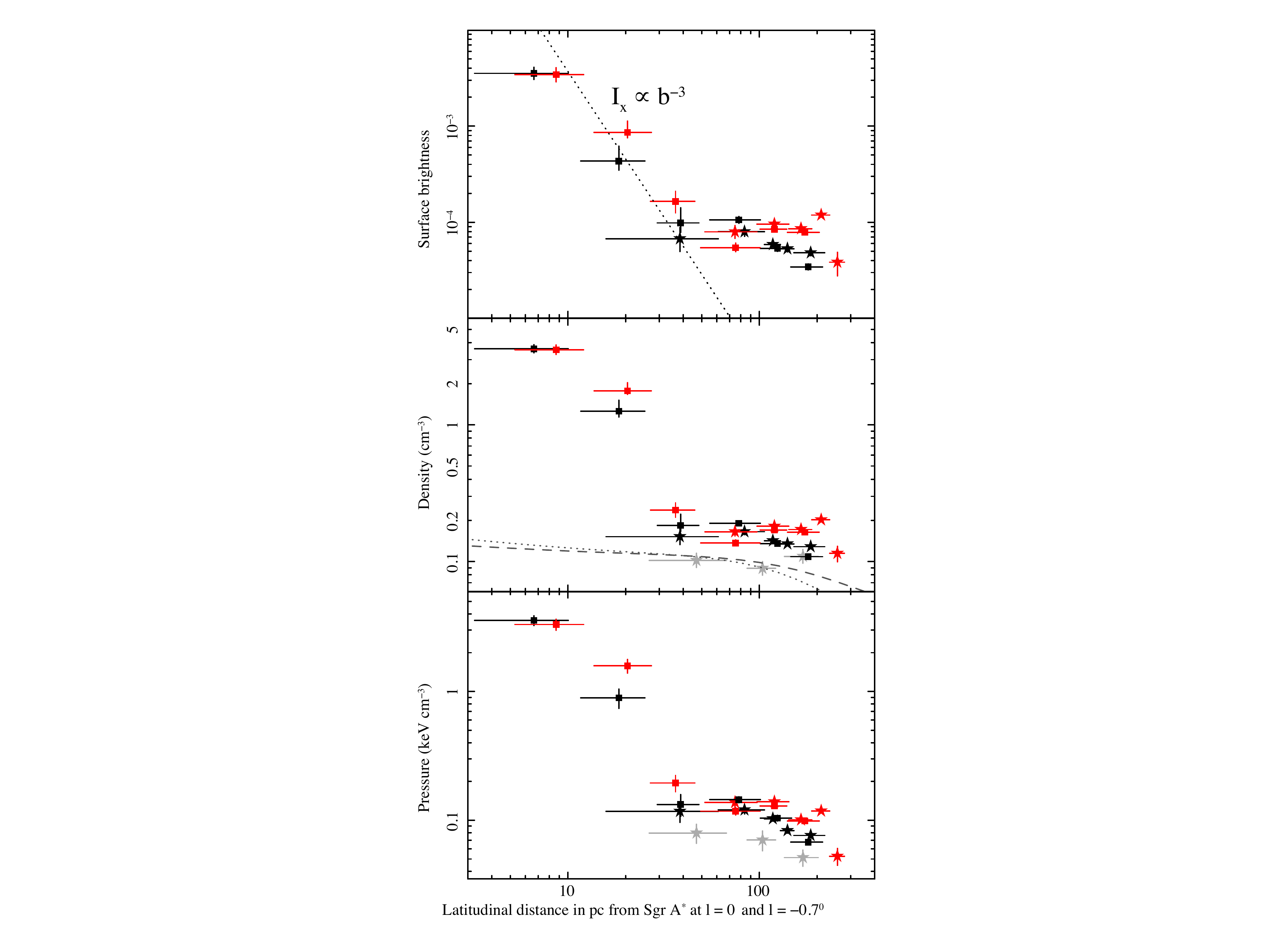}
\vspace{-0.5cm}
\caption{\footnotesize Surface brightness, density and pressure latitudinal profiles of the Galactic centre corona. 
Black and red symbols refer to fits to the data (at $l=0^\circ$) for the northern and southern Chimneys, respectively. 
Similar characteristics are found for the northern and southern Chimneys. 
Stars and squares show results from \xmm\ and \chandra, respectively. 
Grey stars show the latitudinal profile at $l=-0.7^\circ$. 
To estimate densities and pressure, we assumed that the $\pm15$~pc Sgr~A 
lobes have depths of 15~pc, while all other regions have depths of 150~pc 
(e.g., the Chimneys as well as the high latitude emission, etc.). 
Although the $l = -0.7^\circ$ points are at the same Galactic 
latitude as the $l = 0^\circ$ points, they are characterised by a smaller 
density and pressure.
The grey dotted and dashed lines show the density stratification for a 
static/non-rotating and isothermal atmosphere embedded in 
the gravitational potential of the Milky Way with temperature of 
$kT=0.6$ and $1$~keV, respectively. The curves are normalised 
in order to reproduce the densities observed at $l=-0.7^\circ$. 
The densities within the Chimneys are larger by a factor of $\sim2-3$ 
compared with the expectations, while the $\pm15$~pc bipolar lobes 
are more than an order of magnitude larger than expected 
by the Milky Way gravitational potential. Nevertheless, 
the roughly factor of 2 drop in densities within the Chimneys 
is likely related to the drop of Milky Way potential.} 
\label{ProfE}
\end{figure}

\begin{figure}
\addtocounter{figure}{-1}
\centering
\vspace{-2cm}
\includegraphics[width=0.6\textwidth,angle=0,trim=1cm 1cm 1cm 1cm,clip=true]{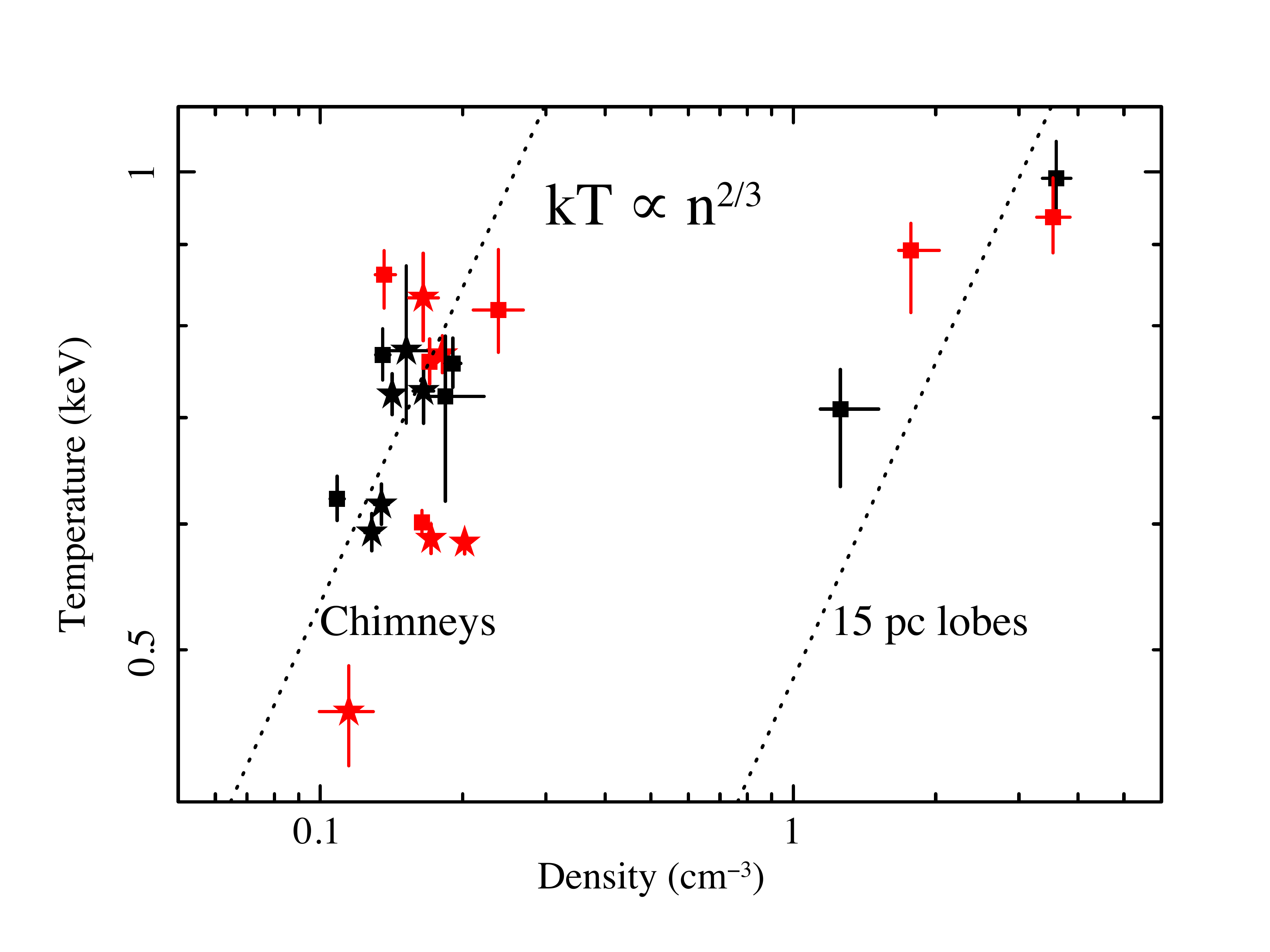}
\vspace{-0.5cm}
\caption{\footnotesize Continued: Temperature vs. density profiles of the Galactic corona. 
Two different adiabatic laws are required to interpolate the $\pm15$~pc
bipolar lobes and the Chimneys. Therefore, if both features are filled 
with warm X-ray emitting plasma, this indicates that the Chimneys and 
the $\pm15$~pc lobes are the manifestation of two distinct outflows.} 
\label{ProfE}
\end{figure}

These $\pm15$~pc bipolar lobes are traced down to parsec scale by plasma 
with temperature $kT\sim0.7-1$ keV. Their X-ray surface brightness declines 
rapidly with latitude $I_x \propto b^{-2}$ to $b^{-3}$ between a few to $\sim15$ pc, 
indicating a strong pressure gradient (see Fig. \ref{Prof}, \ref{ProfE}$^{10,7}$). 
Such a pressure gradient cannot be compensated by the gravitational potential 
of the central region; therefore the gas in the bipolar lobes must be outflowing (Fig. \ref{ProfE}). 
Under the assumption of freely expanding conical or paraboloidal outflow$^{11}$, 
the temperature should drop rapidly with latitude due to adiabatic expansion, 
converting it into a fast and cold gas stream. 
Such scenarios are often invoked for the extra-planar outflows in star-forming 
galaxies$^{12,13,14}$, where the observed X-rays are merely tracers of 
a much more powerful energy- and mass-loaded outflow. 

Alternatively, the observed X-rays could be coming from the shock-heated gas 
that surrounds the lobes, reminiscent of AGN-inflated bubbles in galaxy clusters. 
In this scenario, the temperature of the gas would characterise the expansion 
velocity of the lobes. 

To obtain a conservative limit on the power of the $\pm15$~pc structure, 
we assume that the X-ray gas is volume filling and the outflow velocity is comparable 
to the sound speed, $c_s\approx 500\;{\rm km\;s^{-1}}$, set by the gas temperature. 
The thermal energy of the $\pm15$~pc lobes is then $\sim6\times10^{50}\;{\rm ergs}$ 
and the power of the outflow averaged over the sound-crossing time of the lobes, 
$t_s\sim 3\times10^{4}\;{\rm yr}$, is $L_{15~pc}\sim 8\times10^{38}\;{\rm erg\;s^{-1}}$. 
These are relatively modest requirements in terms of the time-averaged 
energetics that could be satisfied by accretion-powered events from 
the Galactic centre black hole, which may arise e.g., from tidal disruption 
of stars, releasing $10^{51-52}\;{\rm erg}$ every few~$10^{4}\;{\rm yr}$$^{15,16}$ 
and/or by supernovae explosions in the central star cluster$^{17}$.
If the X-ray emitting gas has a very low filling factor, then the actual energetics 
of the $\pm15$~pc outflow could be much larger than the values quoted above. 

At latitudes larger than $\pm15$~pc, the X-ray emission from 
the $\pm15$~pc bipolar lobes blends into the more extended emission 
of the newly discovered Chimneys. Within the northern Chimney, the plasma 
temperature remains roughly constant at $kT\sim0.7-0.8$~keV over its entire 
extent ($\sim160$~pc; Fig. \ref{Prof}). This and the shell-like morphology 
of the northern Chimney are easily understood if this mild outflow encounters 
a cap partly obstructing its further latitudinal expansion$^{6}$. 
The density within the Chimneys gently decreases with latitude 
from $\sim0.2$ at $30$~pc to $\sim0.1\;{\rm cm^{-3}}$ at $160$~pc. 
The thermal energy in the Chimneys, derived from the X-ray data as is done 
above for the $\pm15$~pc lobes, is $\sim 4\times10^{52}$~erg. 
Using again the sound crossing time of the 160~pc structure 
($t_s\sim 3\times10^{5}$~yr) as an estimate for the energy replenishment time, 
one gets the power $L_{160\;{\rm pc}}\sim4\times10^{39}\;{\rm erg\;s^{-1}}$. 
Although this is a factor of 5 higher than the estimated power for the $\pm15$~pc lobes, 
it is still consistent with the supernovae or tidal disruption scenarios. 
Note that for the plasma in the Chimneys, the Galactic gravitational potential may not 
be negligible; therefore the $kT\sim0.7$ keV plasma might be close to hydrostatic 
equilibrium (Fig. \ref{Prof}), rather than outflowing at the sound speed. 
In this scenario, the Chimneys testify the channel excavated by the powerful 
outflows associated with a series of past episodic events and connecting 
the GC with the halo. The past outflows might have been even more collimated 
than its relics (the Chimneys) appear today. 
The long cooling time ($t_{cool}\sim2\times10^7$~yr) and the shallow density 
gradient in the vertical direction do not contradict this scenario.
However, the edge-brightened morphology of the Chimneys rather favours 
the dark (e.g., cold or low density) flow scenario, when X-rays at the edges 
are due to the flow interaction with the denser ISM.

The Chimneys appear to be well confined in the longitudinal direction, 
except for a ``protrusion'' on the eastern side of the northern Chimney (see Methods), 
and they have sharp edges along much of their vertical extent (Fig. \ref{Diff}). 
Given the estimated pressure of the X-ray emitting gas in the Chimneys 
($P\sim 0.1-0.2\;{\rm keV\,cm^{-3}}$) it seems plausible that either dense ISM phases 
or magnetic pressure ($B\sim 60-90\;{\rm \mu G}$ would be required) can provide 
the necessary confinement. For example, radio surveys of the central few degrees 
indicate a rather strong magnetic field$^{18,19,20}$, rising vertically 
from the central molecular zone. 
The almost cylindrical shape of the Chimneys and their widths of $\sim100\;{\rm pc}$ 
at $b>\pm0.2^\circ$, raises the possibility that they are formed by a distributed energy 
injection process in the plane, evidence that would favour a star-formation 
powered scenario. This is corroborated by the observations that the majority 
of the evolved massive stars of the central molecular zone are concentrated 
at the feet of the Chimneys$^{21,22}$. 
It is worth noting that for a flow through a channel with a constant cross section, 
the pressure experiences a modest drop [factor of $\sim 2$] along the cylindrical 
part of the channel (due to the acceleration of the gas and the Galactic 
gravitational potential), but can drop dramatically once the channel flares. 
The observed temperature of the gas in the Chimneys, if viewed as a proxy 
for a typical energy-per-particle value for injected energy and mass, also appears 
to be within a factor of few from what is usually assumed in star-formation driven 
scenarios$^{23}$.

In the direction perpendicular to the Galactic plane, the Chimneys extend beyond 
one degree, where they appear to merge with the FB, as suggested 
by the extrapolation of the boundaries of the extended structures seen in \rosat\ images 
to small latitudes (Fig. \ref{FourPanels}, \ref{Cha}$^{24}$). 
We tentatively associate the transition from the Chimneys to the FB with a modest drop 
of temperature, which in the northern Chimney occurs $\sim$160 pc above the plane. 
In the surface brightness distribution, the transition is also relatively inconspicuous.

With these new data, two questions arise. 
One is whether the inner $\pm15$~pc bipolar lobes and the larger 160~pc Chimneys 
are parts of the same outflow that starts very close to \sgras. 
While this cannot be entirely excluded due to the likely intermittency of the energy 
release, the X-ray data do not directly support the idea that the Chimneys 
are a continuation 
of the inner lobes. For instance, the density-temperature relations shown in Fig. \ref{ProfE} 
indicate that the entropies associated with these two flows are markedly different, 
as shown by the data following two separate adiabatic relations ($kT\propto n^{2/3}$). 
Rather, the inner lobes appear to be embedded into the volume occupied 
by the larger Chimneys. At the same time, the hypothesis that both structures are 
powered by the same physical mechanism cannot be excluded. 
One plausible candidate for such a universal mechanism is the energy released 
by supernovae. In this scenario, it is the strongly uneven distribution of massive 
stars over the GC region (one concentration close to \sgras\ and another one 
distributed over $\pm 50\;{\rm pc}$) that leads to the nested appearance 
of two outflows, rather than to a single one.

The hypothesis in this paper is that the Chimneys could be the channel 
that transports the energy from the active GC region to the FB. 
Their morphology does suggest so. 
The various estimates of the power needed to create and sustain the FB differ by several 
orders of magnitude, from few $10^{40}$ to $10^{44}\;{\rm erg\;s^{-1}}$, 
depending on the assumed source of energy and its intermittency$^{4,25,26}$. 
The lower end of these estimates is within an order of magnitude of 
the Chimney energetic, which is plausibly a lower limit. 
The ``low-power'' models are predominantly associated with quasi-continuous 
star formation wind scenarios$^{25,27,28}$ and the observed morphology 
of the Chimney is reminiscent of these wind models, albeit with some modifications, 
e.g., related with the shape of the ``nozzle'' that confines the flow over $\sim 100$~pc. 
In the supernovae-powered scenario described above, the role of \sgras\ 
(and the $\pm15$~pc bipolar lobes) is subdominant, assuming that the distribution 
of massive stars is a good proxy for the energy release. 
However, extremely energetic events associated with accretion episodes 
onto \sgras, like tidal disruption events and instances of Seyfert-like activity, 
remain a viable scenario too$^{29,30}$. 
In this case, the channel created by the Chimneys could alleviate the propagation 
of matter and energy from \sgras\ to the rarefied and low pressure regions above 
the disc (see, e.g., $^{27}$).

Irrespective of their primary source, the GC Chimneys reveal 
the exchange of energy, plasma and metals between galactic disks 
and their bulges and they provide preferential channels to connect 
galactic centres to their halos. 
Understanding the outflows produced by the low level of accretion 
onto \sgras\ and star-formation at low rates is of central importance 
to learning about processes surrounding low luminosity active 
galactic nuclei and quiescent galaxies.

\newpage 

\appendix

\begin{methods}

\subsection{Data reduction and analysis} 
\label{Analysis}

The basis of our results are smoothed X-ray images in different energy 
bands and spectra of the diffuse X-ray emission of the innermost 
Galactic corona. The X-ray maps are based on two recent \xmm\ large 
observing programs comprising 46 observations of 25~ks exposure 
(see Fig. \ref{Contmap}), 
in addition to all the available archival data (mainly concentrating 
on the scan of the central molecular zone$^{7}$). 
All new observations are performed with the European Photon Imaging 
Camera (EPIC) PN$^{31}$ and MOS CCDs$^{32}$, 
in full-frame CCD readout mode with the medium optical filters applied. 
The \emph{XMM-Newton} data reduction was performed using the Science Analysis System (SAS) version 16.1.0. For more details on the analysis pipeline see $^{33,7}$.
Data products were downloaded from the respective archives, we screened 
times of high background and removed point sources. 
Detector background files were created and subtracted using the task \texttt{evqpb}, 
summing up filter-wheel-closed (FWC) files with 10 times the exposure 
of each observation.

We also used archival observations with the \emph{Chandra} 
Advanced CCD Imaging Spectrometer$^{34}$ imaging (-I) 
CCD array (about 0.1 to 10\,keV energy range). 
The \emph{Chandra} ACIS-I observations were reprocessed using 
the \emph{Chandra} Interactive Analysis of Observations software package$^{35}$ 
version 4.5 and the \emph{Chandra} Calibration Database version 4.5.9. 
For a detailed description of the data reduction pipeline 
see $^{36}$.
We estimated the background in each \emph{Chandra} observation via 
the CIAO tool \texttt{acis\_bkgrnd\_lookup,} which delivers a suitable blank-sky 
background file, provided by the \emph{Chandra} X-ray centre$^{37}$. 
The \emph{Chandra} data have an exposure of about 50\,ks in the central 
scan, 15\,ks in the vertical extensions, and 2\,ks in the large areas at higher 
latitude (see Fig. \ref{Cha}).

The combination of \chandra\ and \xmm\ data provides the highest 
spatial ($\mathrm{\sim1~arcsec}$) and spectral resolution ($\mathrm{\sim100~eV}$ 
full width half maximum, FWHM) available in soft X-rays.

For the \xmm\ data we performed source detection on each individual 
observation and all detected point sources have been masked out. 
Readout streaks were included in background images. 
For \chandra\ data, we performed source detection on all datasets, 
simultaneously in multiple energy bands. A second point source detection 
was run on the merged images and, after carefully screening the detection list, 
the point sources were masked out from the image (in some cases 
additional regions were masked to screen dust-scattering halos and 
readout streaks of very bright sources). 
We then computed the background subtracted, exposure-corrected 
and adaptively smoothed images, resulting into smoothed flux 
images of diffuse emission (foreground, unresolved point sources, 
and intrinsic GC hot plasma emission).

The flux images were then binned into regions of equal 
signal-to-noise (S/N) ratio 50 ($\rm{>2.5\times10^3}$ counts per bin) 
for \emph{Chandra} and S/N 100 ($\rm{>1\times10^4}$ counts per bin) 
for \emph{XMM-Newton}. 
The difference was chosen due to more stable background in \emph{Chandra} 
compared to \emph{XMM-Newton}. 
The contour binning technique \texttt{contbin}$^{40}$ was used for this task. 
For each observation overlapping with one of the bins, spectra and response 
files were created and all contributions were added with proper weighting 
to obtain one spectrum for each spatial bin. 
These small scale bins were then added up to obtain higher S/N spectra 
for regions of interest (see Figs. \ref{RegCha} and \ref{Regh}). 
Each of the regions shown in Figs. \ref{RegCha} and \ref{Regh} 
consists of 5-7 merged initial bins.
The spectra in the 0.5-7 keV energy range were then fitted with emission 
models using \texttt{XSPEC}$^{41}$ version 12.9.1n. 
The best fit parameters were used to derive the profiles shown 
in Figs. \ref{Prof}, \ref{ProfE}, and \ref{Profh}. 

We identified the following \texttt{XSPEC} model as best general description 
of the data (unless stated otherwise for tests), 
{\sc Tbabs (apec+pow+gauss+gauss)+Tbabs  apec }
where the first foreground absorption (\texttt{Tbabs} model) is the absorption 
towards the GC, the first \texttt{apec} model is the intrinsic 
$\mathrm{\sim1~keV}$ plasma at the GC, the \texttt{pow} and 
the two \texttt{gauss} models describe the powerlaw and two narrow Gaussian 
lines at 6.4 and 6.7\,keV caused by unresolved point sources around 
the GC and non-thermal emission$^1$ (the photon index of the power law 
was fixed to $\Gamma=1.8$). 
The second \texttt{apec} model with lower foreground 
absorption ($\mathrm{0.7\times10^{22}~cm^{-2}}$) and a temperature 
of about 0.2\,keV describes a "foreground" emission (i.e., between 
the observer and the GC, e.g., such as the contribution from the local 
bubble) that was observationally chosen to fit the data.
Throughout this work we used the \texttt{apec} model for collisionally-ionised 
diffuse gas, which is based on the ATOMDB code v3.0.7$^{42}$.
Uncertainties are quoted on the $\mathrm{1\sigma}$ level unless stated otherwise. 
Abundances are according to Solar abundances as from $^{43}$. 
The distance to the GC was assumed to be 8\,kpc$^{44}$. 

The profiles shown in Figs. \ref{Prof}, \ref{ProfE}, and \ref{Profh} are derived 
from the best fit values of the \texttt{apec} model to the intrinsic $\mathrm{\sim1~keV}$
plasma component at the GC.
The free fit parameters are the plasma temperature ($\mathrm{T~[keV]}$)
and the model normalization ($\mathrm{\eta~[cm^{-5}]}$). The surface
brightness profiles are given as $\eta~arcmin^{-2}$.
Density ($\mathrm{n~[cm^{-3}]}$) and pressure
($\mathrm{P~[keV~cm^{-3}]}$) were derived assuming full ionization with
10 per cent helium and 90 per cent hydrogen (i.e. $\mathrm{n_e \sim
1.2n_p}$),

\begin{equation}
\mathrm{n = \sqrt{\eta \times \frac{19.3~\pi~D_A^2}{~V}}}
\end{equation}

\begin{equation}
\mathrm{P = T \times n}
\end{equation}
where $\mathrm{D_A}$ is the distance to the source in cm and V is the
assumed volume in $\mathrm{cm^3}$. The volume was calculated as the
extraction surface area times the assumed depth. For the $\mathrm{\pm15~pc}$
lobes we assumed a constant depth of $\mathrm{15~pc}$.
For the remaining diffuse emission, we assumed a uniform depth of 
$\mathrm{150~pc}$, as suggested by the size of the Chimneys. 
A caveat of this approach is that the density beyond the Chimneys, 
at the foot of the Fermi-Bubbles, might be overestimated by a large 
factor, because that emission probably comes from a much larger
volume. Indeed, if we assume a depth of 1.5\,kpc for the high latitude 
emission, the expected particle density would drop 
to $\mathrm{\sim10^{-2}~cm^{-3}}$.

Comparing the surface brightness of the Fermi-bubble X-ray counterparts 
in \emph{ROSAT} images with the high latitude emission observed 
with \emph{XMM-Newton}, we found good agreement. 
The photon count rate in a \emph{ROSAT} 1-2\,keV (assuming $\sim60~cm^2$ 
effective area) mosaic image (Fig. \ref{FourPanels}) and in the \emph{XMM-Newton} 
image ($\sim1500~cm^2$) of the same band are consistent 
at $\mathrm{\sim3\times10^{-6}~s^{-1}~cm^{-2}~arcmin^{-2}}$ (1-2\,keV).

\subsection{Other sources powering Sgr\,A's bipolar lobes}

In the main article we suggested that Sgr\,A's bipolar lobes are 
primarily powered by supernovae explosions and tidal disruption 
events. However, this does not exclude that other processes might 
provide a contribution in powering such features. 

For example, can the current activity of \sgras\ inflate the $\pm15$~pc 
Sgr\,A's bipolar lobes$^{30}$? 
Studies of clusters of galaxies demonstrate that supermassive black holes 
accreting at low Eddington rates do generate outflows that inflate bubbles 
into the intra-cluster medium and have kinetic luminosities orders of magnitude 
larger than the instantaneous radiative power of these BH$^{45,46}$. 
For \sgras\ the quiescent radiative luminosity $\sim10^{36}\;{\rm erg\;s^{-1}}$, 
when expressed in Eddington units, is even lower than for the black holes 
in galaxy clusters, firmly placing this object deep into a very low-luminosity 
regime$^{47}$. 
It is plausible that in this regime the outflow power can exceed the radiative 
luminosity by a large factor, opening a possibility that \sgras\ could provide 
the mechanical luminosity, needed to power the $\pm15$ pc Sgr\,A bipolar 
lobes, even without invoking additional tidal disruption-like major outbursts.  

\subsection{Uncertainty related with the unknown filling factor 
and abundances}
\label{filling}

The X-ray morphologies of all structures seen in \rosat, 
\xmm\ and \chandra\ images are very complicated with signs 
of edge-brightening. 
It is plausible that the X-ray-emitting gas is not volume filling, 
but represents patches of denser gas interacting with an outflow, 
as suggested by many simulations that involve rapid expansion 
and adiabatic cooling of matter in the outflow$^{11}$.  
We nevertheless assume in the main paper the volume filling 
factor $f=1$, to derive the gas density in the observed structures 
and estimate their total energy content. 
For $f<1$, the recovered density scales as $n_{obs}\propto n_{true}f^{1/2}$. 
Assuming that the energy density is uniform over the structure 
(in the longitudinal direction), the estimated total energy will be also 
lower than the true one by a factor $f^{1/2}$ and therefore should 
be treated as a lower limit. 

Throughout the paper, we assume Solar abundances of the X-ray 
emitting plasma. This assumption is performed in order to avoid spurious 
degeneracies between the fitted parameters. We note that the X-ray 
emitting plasma in the plane, within the central molecular zone, has Solar 
or slightly super-Solar metallicities$^{10,7,48}$. 
The metallicity is expected to drop with latitude. Indeed, 
the halo metallicity is expected to be significantly lower than Solar, 
with metallicities of the X-ray emitting plasma of the order of $\sim0.3$ 
Solar, inside the FB$^{9,26}$. 
To quantify the consequences of this assumption, we fitted one of the 
\xmm\ spectra of the northern Chimney changing the assumed metallicity 
from $1.5$ to $0.3$ Solar. We observed that the best fit temperature and 
emissivity changed by $\sim5$ and $\sim10$ \%, respectively. 

\begin{figure}
\centering
\vspace{-1cm}
\includegraphics[width=0.9\textwidth,angle=0,trim=0cm 2cm 0cm 2cm,clip=true]{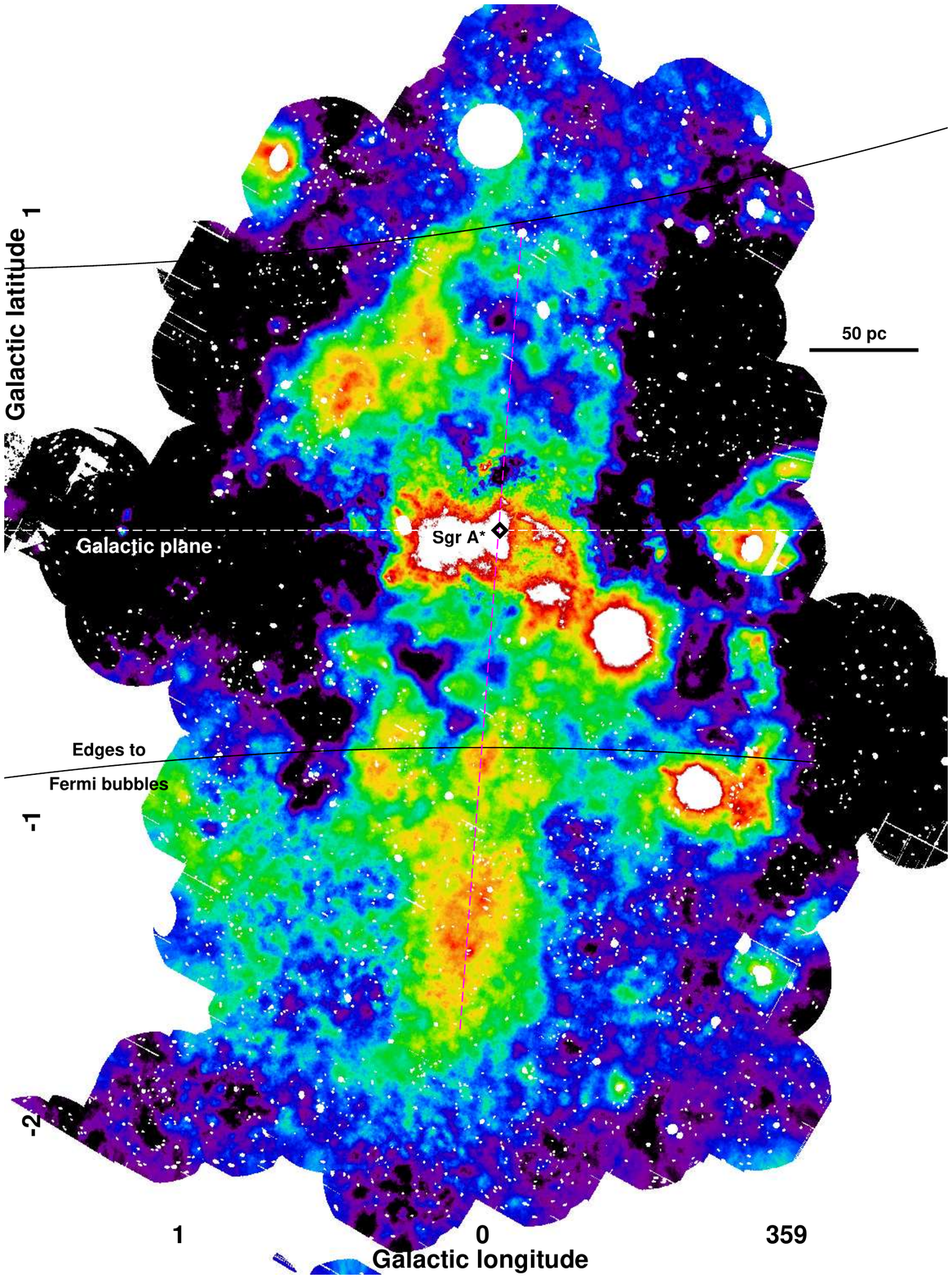}
\caption{\footnotesize X-ray emission from the central degrees of the Milky Way. 
Bright X-ray emission traces the coherent edge brightened 
shell-like feature, dubbed northern Chimney, located north of \sgras\ and 
characterised by a diameter of $\sim160$~pc. On the opposite side, 
the southern Chimney appears as a bright linear feature. 
Bright X-ray emission is observed at high latitude 
($|b|\gsimeq1^\circ$), corresponding to the X-ray counterparts of the \fermi\ 
bubble. The magenta dashed line intersects both Chimneys, 
passing through \sgras. 
The map shows the X-ray emissivity within the 1.5-2.6 keV energy band. 
Point sources have been removed. Larger circles have been excised 
to remove the dust scattering haloes around bright sources. }
\label{Contmap}
\end{figure}

\begin{figure}
\centering
\vspace{-2cm}
\includegraphics[width=0.9\textwidth,angle=0,trim=0cm 0cm 0cm 0cm,clip=true]{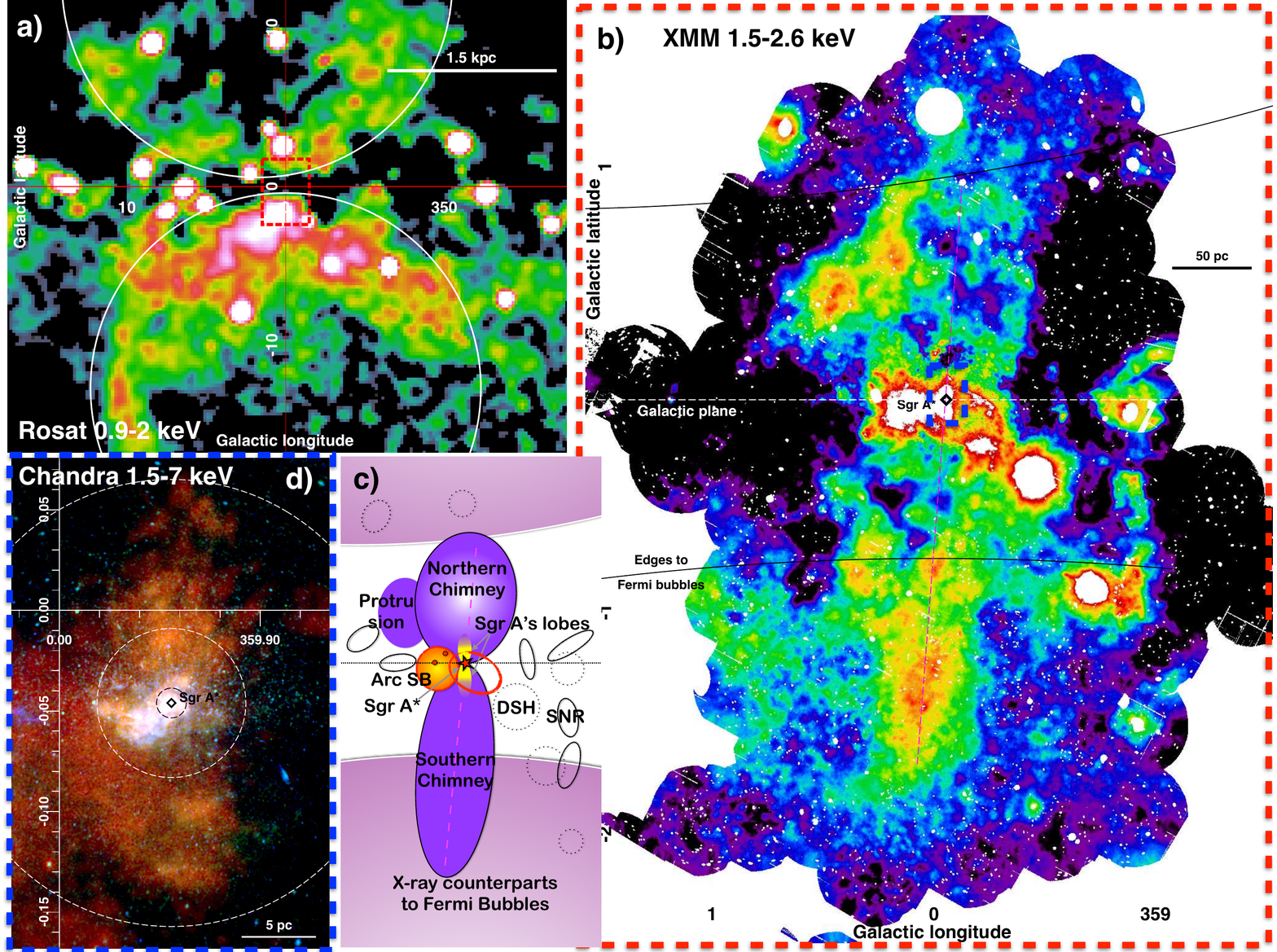}
\caption{\footnotesize {\it (Panel a)} \rosat\ (0.9-2~keV energy band) large scale map of the GC. 
The X-ray counterparts of the \fermi\ bubbles are strong X-ray emitters.  
The edges (white ellipses) are clearly detected on scales of several degrees, 
while they become confused (because of the short exposure 
of 200-300~s and soft X-ray energy band) close to the plane. 
The red dashed line indicates the \xmm\ area covered by our survey.
{\it (Panel b)} \xmm\ map zooming into the central degrees of the Milky Way. 
The magenta dashed line intersects both Chimneys, 
passing through \sgras. 
The map shows the X-ray emissivity within the 1.5-2.6 keV energy band (see Fig. \ref{Contmap}). 
{\it (Panel c)}
Schematic view of the main diffuse X-ray emitting features 
within the central $\sim500$~pc from \sgras. 
The red star and the yellow ellipses indicate the position of \sgras\ 
and of Sgr~A's $\pm15$~pc bipolar lobes. 
The large violet ellipses indicate the location and extension of 
the X-ray counterpart of the GCL with shell like morphology (northern
Chimney), of its eastern protrusion and of its roughly symmetric southern 
counterpart (southern Chimney). The orange filled ellipse, the two red circles 
and the red ellipse indicate the location of the Arc super-bubble, 
the Quintuplet and Arches clusters and the super-bubble candidate 
G359.77-0.09$^7$. 
The pink regions indicate the location of the edges to the \fermi\ bubbles. 
The dotted circles and solid ellipses indicate 
the position of bright X-ray sources with intense dust scattering 
halos and known supernova remnants. 
{\it (Panel d)} \chandra\ RGB map zooming into the central tens of parsecs 
of the Galaxy. The $\pm15$~pc lobes are clearly visible in orange colour. 
The dashed circles have radii of 1, 5, and 15 pc.}
\label{FourPanels}
\end{figure}

\begin{figure}
\centering
\includegraphics[width=0.9\textwidth,angle=0,trim=1cm 1cm 1cm 3cm,clip=true]{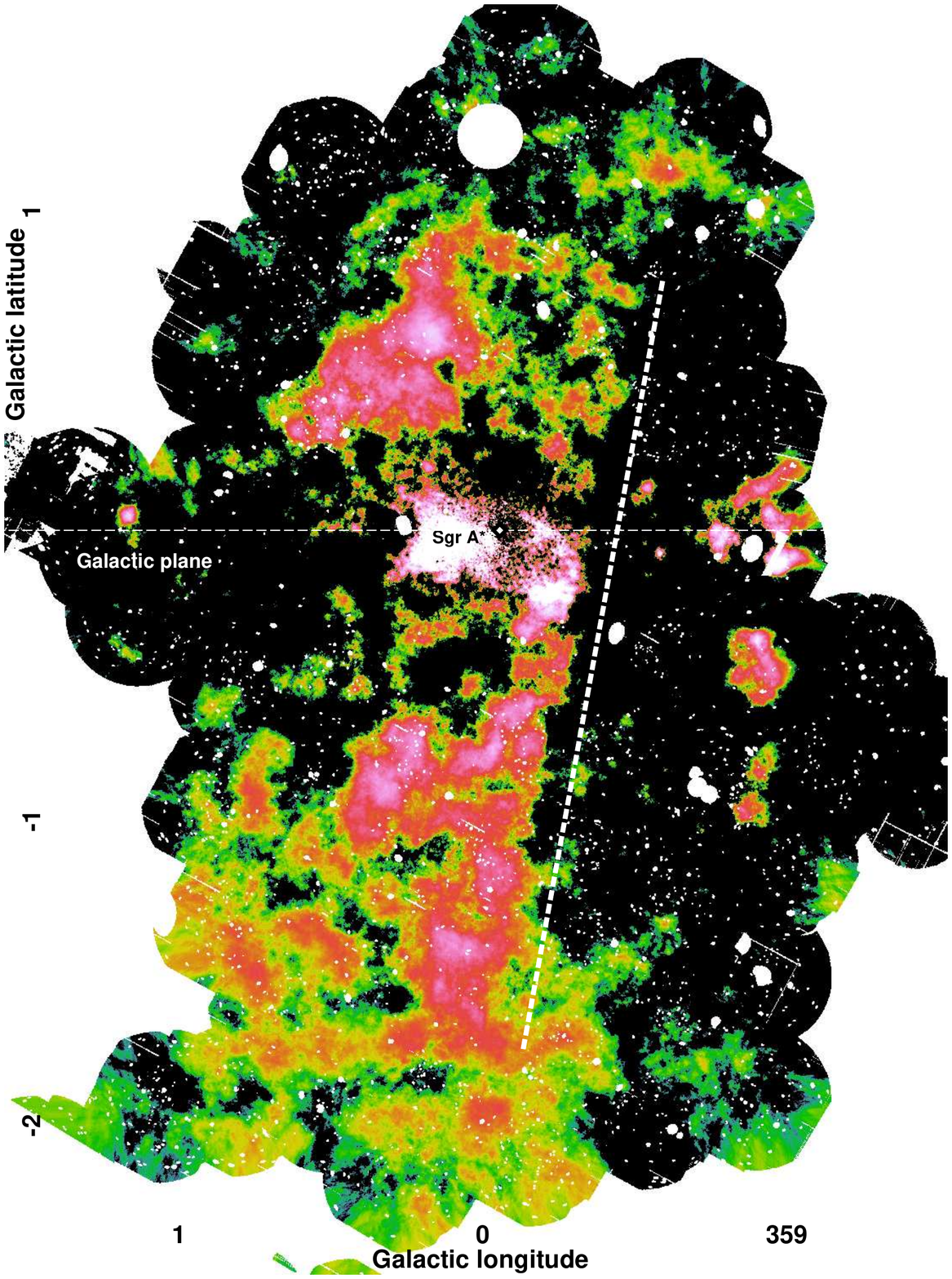}
\caption{\footnotesize Continuum subtracted {\sc S xv} line emission. 
The map highlights sources characterised by bright soft X-ray 
emission lines, such as diffuse thermally emitting plasma. 
On the other hand, point sources and dust scattering halos 
are efficiently removed. 
The white dashed tilted line indicates a linear ridge, $\sim450$~pc 
long, that appears to run west of the Chimneys. }
\label{Diff}
\end{figure}

\begin{figure}
\centering
\includegraphics[width=\textwidth,angle=0,trim=0cm 5cm 0cm 5cm,clip=true]{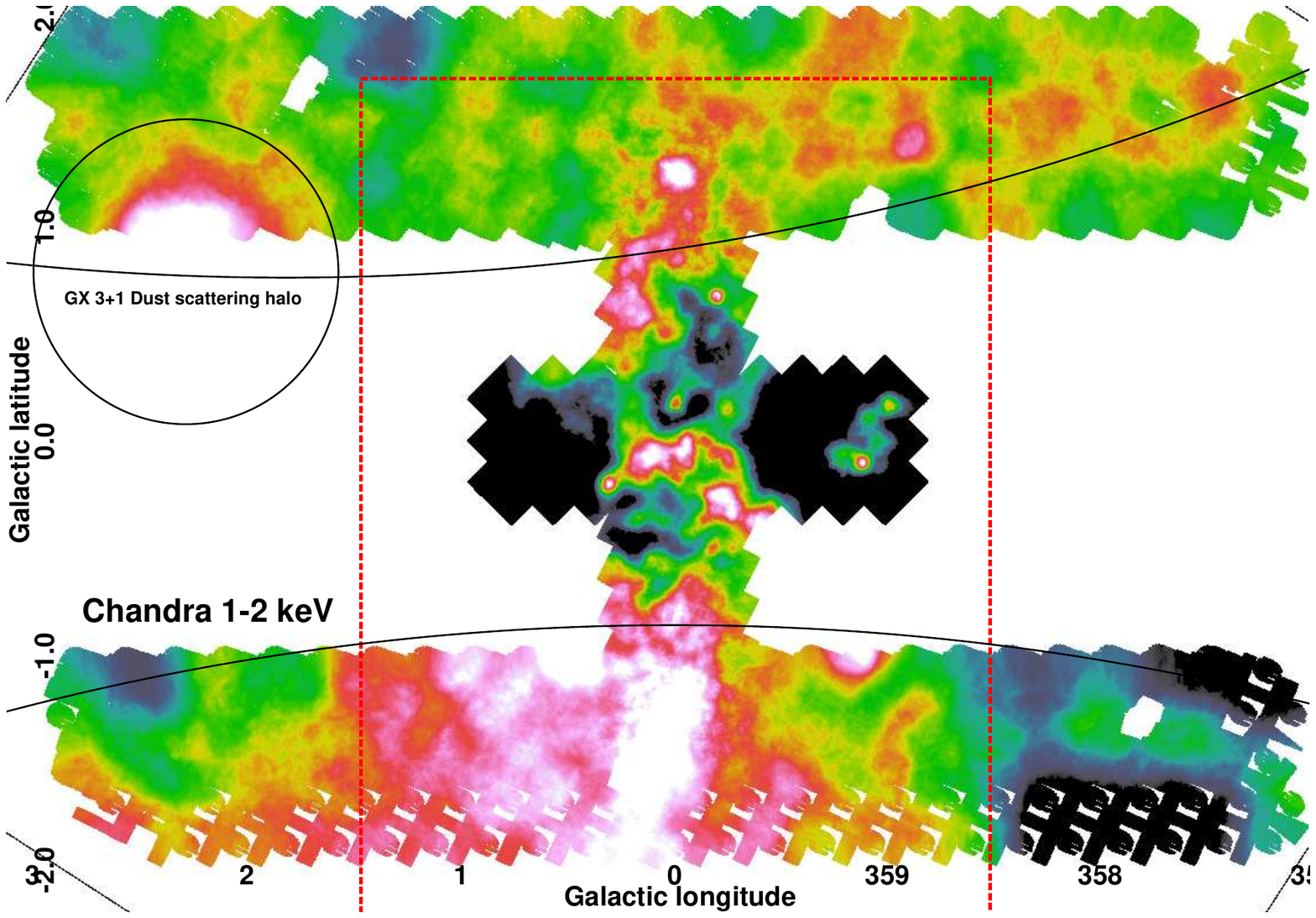}
\caption{\footnotesize Chandra 1-2~keV map of the Galactic centre. 
The same structures seen in the \xmm\ 1.5-2.6 keV image (Fig. \ref{Contmap}, footprint indicated by red dashed box) are observed also in the shallower \chandra\ image. 
Despite the 1-2~keV band map is more affected by neutral 
absorption towards the plane, compared with the 1.5-2.6 keV
band (Fig. \ref{Contmap}), we display it here in order to emphasise 
the residual high latitude emission (e.g., inside the \fermi\ bubbles). 
}
\label{Cha}
\end{figure}

\begin{figure}
\centering
\includegraphics[width=0.7\textwidth,angle=0,trim=5cm 0cm 5cm 0cm,clip=true]{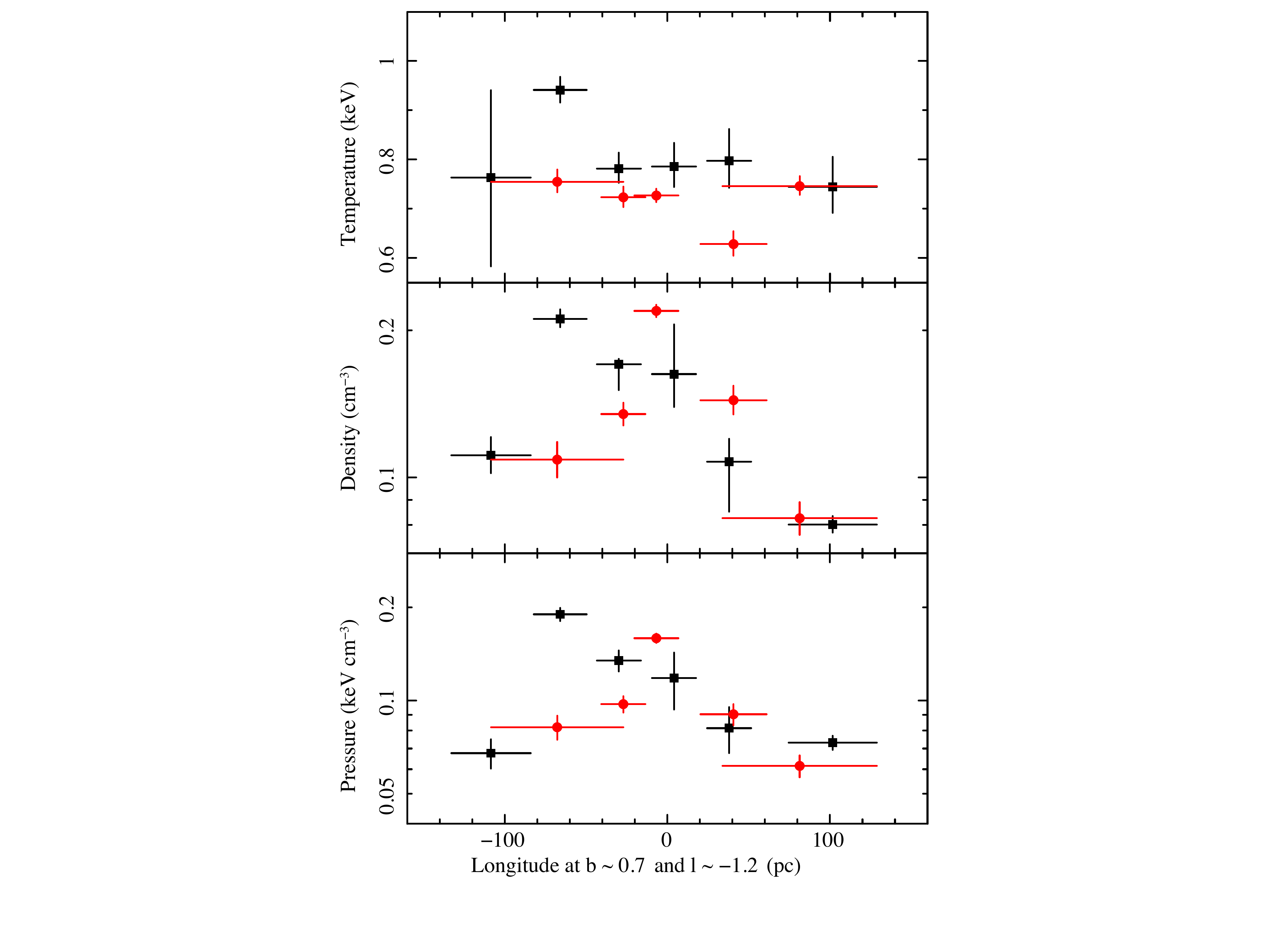} 
\caption{\footnotesize From top to bottom: temperature; density and pressure 
profiles as a function of longitude cutting through the northern and southern
Chimneys at latitudes $b\sim0.7^\circ$ and $b\sim-1.2^\circ$
(see Fig. \ref{Regh}). Positive values indicate Galactic West 
(West is right as in an image in Galactic coordinates). 
Black squares and red circles indicate the longitudinal 
cut through the northern and southern Chimneys, respectively. 
The densities have been estimated assuming a volume $V=PA^{3/2}$, 
where PA stands for the projected area. } 
\label{Profh}
\end{figure}

\begin{figure}
\centering
\vspace{-2cm}
\includegraphics[width=0.9\textwidth,angle=0,trim=8cm 1cm 8cm 1cm,clip=true]{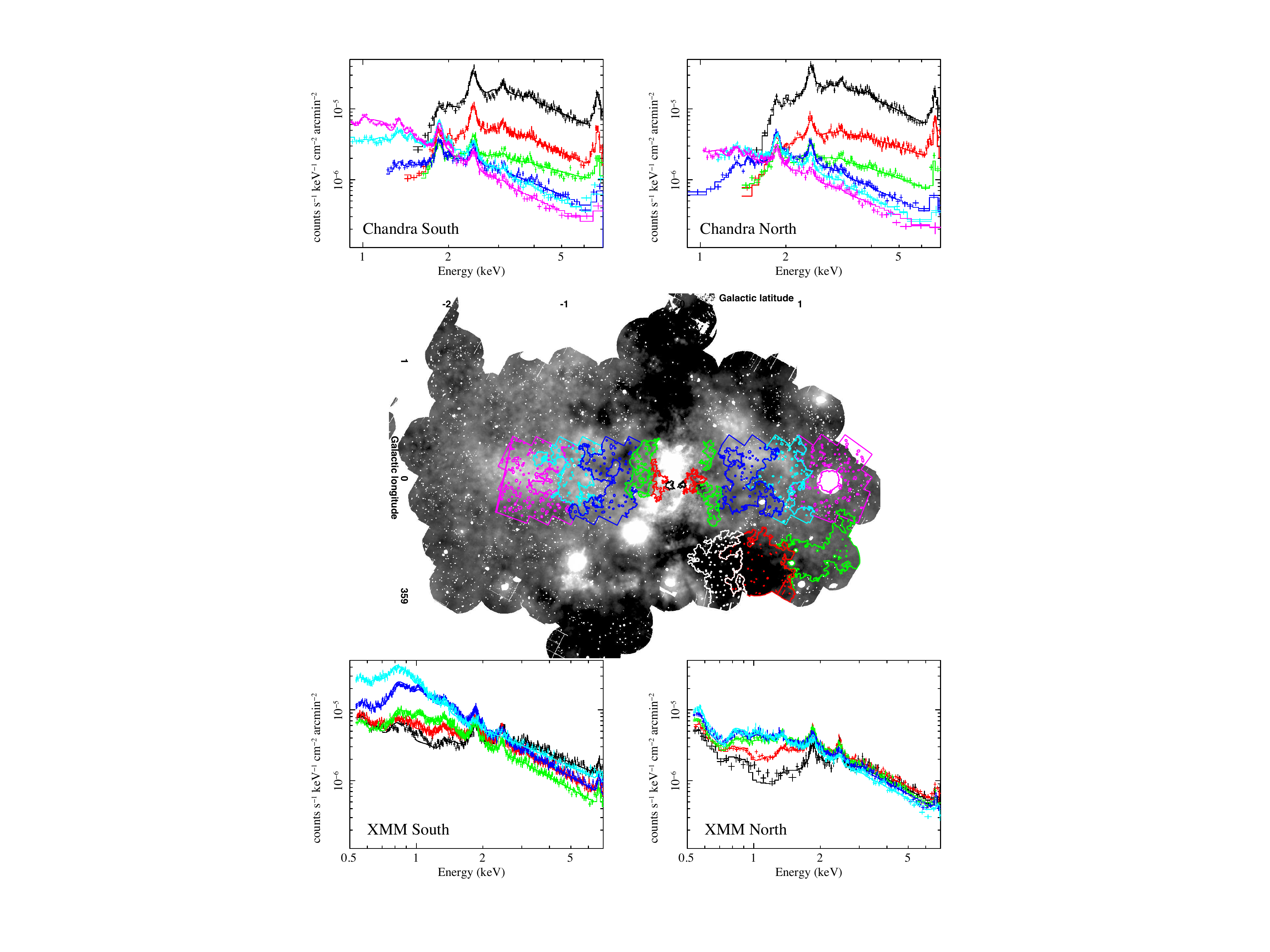}
\vspace{-1.5cm}
\caption{\footnotesize {\it (Top panels)} \chandra\ spectra used to derive the latitudinal 
profiles at $l=0^\circ$ and shown in Fig. \ref{Prof} for the southern 
(left) and northern (right) sides, respectively. 
{\it (Central panel)} \xmm\ 1.5-2.6~keV emission map in greyscale, 
showing the extraction areas of the \chandra\ spectra used to derive 
the latitudinal profiles at $l=0^\circ$, with colours corresponding to those 
of the spectra. 
The areas at $l\sim-0.7^\circ$ indicate the extraction regions of the 
spectra used to derive the grey constraints in Fig. \ref{Prof}. 
{\it (Bottom panel)} \xmm\ spectra used to derive the latitudinal profiles of the diffuse 
emission in Fig. \ref{Prof}.} 
\label{RegCha}
\end{figure}

\begin{figure}
\vspace{-2cm}
\includegraphics[width=\textwidth,angle=0,trim=0cm 1cm 0cm 1cm,clip=true]{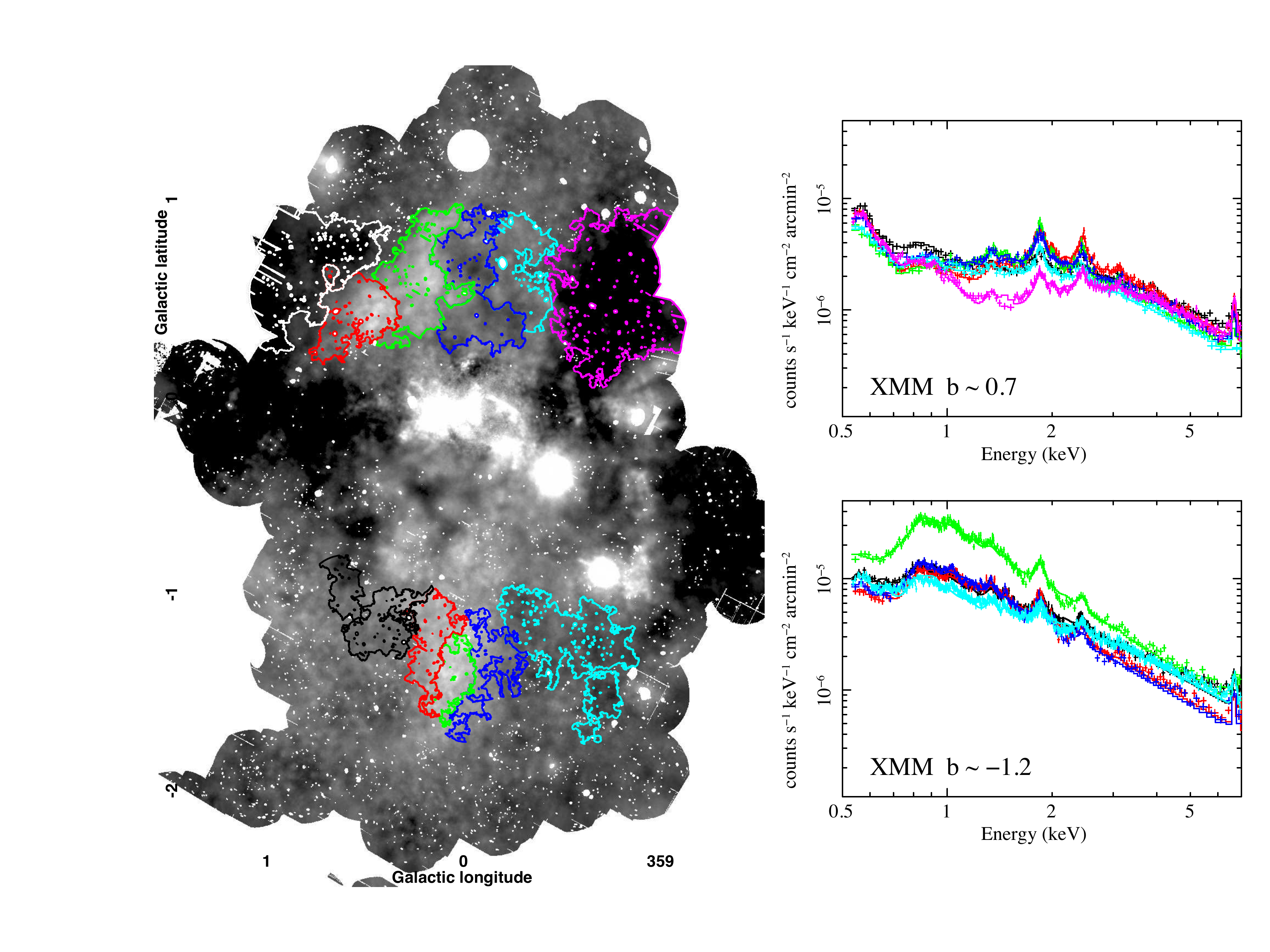}
\caption{\footnotesize {\it (Top panels)} \xmm\ spectra used to measure the longitudinal 
profiles shown in Fig. \ref{Profh}. 
{\it (Bottom panel)} \xmm\ 1.5-2.6~keV map displaying the regions used 
to extract the spectra corresponding to a longitudinal 
cut through the northern shell (at $b\sim0.7^\circ$) and through 
the jet-like feature (at $b\sim-1.2^\circ$). } 
\label{Regh}
\end{figure}

\end{methods}

\newpage

\begin{addendum}
 \item This work is dedicated to Alice Maria Ponti. GP acknowledges 
Andrea Merloni for discussion and the Max-Planck-Institut f{\"u}r 
Astronomie Heidelberg for hospitality. 
This research has made use both of data obtained with XMM-Newton, 
an ESA science mission with instruments and contributions directly 
funded by ESA Member States and NASA and with the Chandra Data
Archive and software provided by the Chandra X-ray Center (CXC).
G. Ponti and F. Hofmann acknowledge financial support from the Bundesministerium 
f\"{u}r Wirtschaft und Technologie/Deutsches Zentrum f\"{u}r Luft- 
und Raumfahrt (BMWI/DLR, FKZ 50 OR 1604, FKZ 50 OR 1715
and FKZ 50 OR 1812) and the Max Planck Society. AG and RT 
acknowledge support from CNES.
 \item[Competing Interests] The authors declare that they have no
competing financial interests.
 \item[Data and Code Availability] The datasets analysed during the current study 
 and the software to perform the analysis are available in the \xmm, \chandra\ and \rosat\ 
 repository: http://nxsa.esac.esa.int, https://www.cosmos.esa.int, 
 http://cxc.harvard.edu and http://www.xray.mpe.mpg.de/rosat/archive.
 \item[Correspondence] Correspondence and requests for materials should be addressed to\\ G. Ponti (ponti@mpe.mpg.de).
\end{addendum}


\begin{thebibliography}{1}

\bibitem{2013ASSP...34..331P} Ponti, G., Morris, M.~R., Terrier, R., \& Goldwurm, A.\ Traces of Past Activity in the Galactic Centre. Cosmic Rays in Star-Forming Environments, 34, 331 (2013)

\bibitem{2003ANS...324..167M} Morris, M., Baganoff, F., Muno, M., et al.\ Deep X-Ray Imaging of the Central 20 Parsecs of the Galaxy with Chandra. Astronomische Nachrichten Supplement, 324, 167 (2003)

\bibitem{2016ApJ...817..171Z} Zhao, J.-H., Morris, M.~R., \& Goss, W.~M.\ A New Perspective of the Radio Bright Zone at The Galactic Center: Feedback from Nuclear Activities. \apj, 817, 171 (2016)

\bibitem{2010ApJ...724.1044S} Su, M., Slatyer, T.~R., \& Finkbeiner, D.~P.\ Giant Gamma-ray Bubbles from Fermi-LAT: Active Galactic Nucleus Activity or Bipolar Galactic Wind? \apj, 724, 1044 (2010)

\bibitem{1984Natur.310..568S} Sofue, Y., \& Handa, T.\ A radio lobe over the galactic centre. \nat, 310, 568 (1984)

\bibitem{2010ApJ...708..474L} Law, C.~J.\ A Multiwavelength View of a Mass Outflow from the Galactic Center. 
\apj, 708, 474 (2010)

\bibitem{2015MNRAS.453..172P} Ponti, G., Morris, M.~R., Terrier, R., et al.\ The XMM-Newton view of the central degrees of the Milky Way. \mnras, 453, 172 (2015)

\bibitem{2010RvMP...82.3121G} Genzel, R., Eisenhauer, F., \& Gillessen, S.\ The Galactic Centre massive black hole and nuclear star cluster. Reviews of Modern Physics, 82, 3121 (2010) 

\bibitem{2013ApJ...779...57K} Kataoka, J., Tahara, M., Totani, T., et al.\ Suzaku Observations of the Diffuse X-Ray Emission across the Fermi Bubbles' Edges. \apj, 779, 57 (2013) 

\bibitem{2013MNRAS.434.1339H} Heard, V., \& Warwick, R.~S.\ XMM-Newton observations of the Galactic Centre Region - I. The distribution of low-luminosity X-ray sources. \mnras, 434, 1339 (2013) 

\bibitem{1985Natur.317...44C} Chevalier, R.~A., \& Clegg, A.~W.\ Wind from a starburst galaxy nucleus. \nat, 317, 44 (1985) 

\bibitem{1990ApJS...74..833H} Heckman, T.~M., Armus, L., \& Miley, G.~K.\ On the nature and implications of starburst-driven galactic superwinds. \apjs, 74, 833 (1990) 

\bibitem{1994ApJ...430..511S} Suchkov, A.~A., Balsara, D.~S., Heckman, T.~M., \& Leitherer, C.\ Dynamics and X-ray emission of a galactic superwind interacting with disk and halo gas. \apj, 430, 511 (1994) 

\bibitem{2017MNRAS.466.1213K} Krumholz, M.~R., Kruijssen, J.~M.~D., \& Crocker, R.~M.\ A dynamical model for gas flows, star formation and nuclear winds in galactic centres. \mnras, 466, 1213 (2017) 

\bibitem{1988Natur.333..523R} Rees, M.~J.\ Tidal disruption of stars by black holes of 10 to the 6th-10 to the 8th solar masses in nearby galaxies. \nat, 333, 523 (1988) 

\bibitem{2018MNRAS.478.4030G} Generozov, A., Stone, N.~C., Metzger, B.~D., \& Ostriker, J.~P.\ Circumnuclear media of quiescent supermassive black holes. \mnras, 478, 4030 (2018) 

\bibitem{2013ApJ...764..155L} Lu, J.~R., Do, T., Ghez, A.~M., et al.\ Stellar Populations in the Central 0.5 pc of the Galaxy. II. The Initial Mass Function. \apj, 764, 155 (2013) 

\bibitem{2006JPhCS..54....1M} Morris, M.\ The Galactic Center Magnetosphere. Journal of Physics Conference Series, 54, 1 (2006)

\bibitem{2010Natur.463...65C} Crocker, R.~M., Jones, D.~I., Melia, F., Ott, J., \& Protheroe, R.~J.\ A lower limit of 50 microgauss for the magnetic field near the Galactic Centre. \nat, 463, 65 (2010) 

\bibitem{2013Natur.501..391E} Eatough, R.~P., Falcke, H., Karuppusamy, R., et al.\ A strong magnetic field around the supermassive black hole at the centre of the Galaxy. \nat, 501, 391 (2013) 

\bibitem{2010ApJ...710..706M} Mauerhan, J.~C., Muno, M.~P., Morris, M.~R., Stolovy, S.~R., \& Cotera, A.\ Near-infrared Counterparts to Chandra X-ray Sources Toward the Galactic Center. II. Discovery of Wolf-Rayet Stars and O Supergiants. \apj, 710, 706 (2010)

\bibitem{2012MNRAS.425..884D} Dong, H., Wang, Q.~D., \& Morris, M.~R.\ A multiwavelength study of evolved massive stars in the Galactic Centre. \mnras, 425, 884 (2012)

\bibitem{2000MNRAS.314..511S} Strickland, D.~K., \& Stevens, I.~R.\ Starburst-driven galactic winds - I. Energetics and intrinsic X-ray emission. \mnras, 314, 511 (2000)

\bibitem{2003ApJ...582..246B} Bland-Hawthorn, J., \& Cohen, M.\ The Large-Scale Bipolar Wind in the Galactic Center. \apj, 582, 246 (2003)

\bibitem{2011PhRvL.106j1102C} Crocker, R.~M., \& Aharonian, F.\ Fermi Bubbles: Giant, Multibillion-Year-Old Reservoirs of Galactic Center Cosmic Rays. Physical Review Letters, 106, 101102 (2011)

\bibitem{2018Galax...6...27K} Kataoka, J., Sofue, Y., Inoue, Y., et al.\ X-Ray and Gamma-Ray Observations of the Fermi Bubbles and NPS/Loop I Structures. Galaxies, 6, 27 (2018)

\bibitem{2014MNRAS.444L..39L} Lacki, B.~C.\ The Fermi bubbles as starburst wind termination shocks. \mnras, 444, L39 (2014)

\bibitem{2017MNRAS.470.1982S} Sofue, Y.\ The 200-pc molecular cylinder in the Galactic Centre. \mnras, 470, 1982 (2017)

\bibitem{2013ApJ...778...58B} Bland-Hawthorn, J., Maloney, P.~R., Sutherland, R.~S., \& Madsen, G.~J.\ Fossil Imprint of a Powerful Flare at the Galactic Center along the Magellanic Stream. \apj, 778, 58 (2013)

\bibitem{2017MNRAS.466.1477R} Roberts, S.~R., Jiang, Y.-F., Wang, Q.~D., \& Ostriker, J.~P.\ Towards self-consistent modelling of the Sgr A* accretion flow: linking theory and observation. \mnras, 466, 1477 (2017)

\end{thebibliography}

\newpage

\begin{thebibliography}{2}

\bibitem[31]{2001A&A...365L..18S} Str{\"u}der, L., Briel, U., Dennerl, K., et al.\ The European Photon Imaging Camera on XMM-Newton: The pn-CCD camera. \aap, 365, L18 (2001) 

\bibitem[32]{2001A&A...365L..27T} Turner, M.~J.~L., Abbey, A., Arnaud, M., et al.\ The European Photon Imaging Camera on XMM-Newton: The MOS cameras : The MOS cameras. \aap, 365, L27 (2001)

\bibitem[33]{2012A&A...545A.128H} Haberl, F., Sturm, R., Ballet, J., et al.\ The XMM-Newton survey of the Small Magellanic Cloud. \aap, 545, A128 (2012)

\bibitem[34]{2003SPIE.4851...28G} Garmire, G.~P., Bautz, M.~W., Ford, P.~G., Nousek, J.~A., \& Ricker, G.~R., Jr.\ Advanced CCD imaging spectrometer (ACIS) instrument on the Chandra X-ray Observatory. \procspie, 4851, 28 (2003)

\bibitem[35]{2006SPIE.6270E..1VF} Fruscione, A., McDowell, J.~C., Allen, G.~E., et al.\ CIAO: Chandra's data analysis system. \procspie, 6270, 62701V (2006)

\bibitem[36]{2016A&A...585A.130H} Hofmann, F., Sanders, J.~S., Nandra, K., Clerc, N., \& Gaspari, M.\ Thermodynamic perturbations in the X-ray halo of 33 clusters of galaxies observed with Chandra ACIS. \aap, 585, A130 (2016)

\bibitem[37]{2003ApJ...583...70M} Markevitch, M., Bautz, M.~W., Biller, B., et al.\ Chandra Spectra of the Soft X-Ray Diffuse Background. \apj, 583, 70 (2003)

\bibitem[38]{2006MNRAS.371..829S} Sanders, J.~S.\ Contour binning: a new technique for spatially resolved X-ray spectroscopy applied to Cassiopeia A. \mnras, 371, 829 (2006)

\bibitem[39]{1996ASPC..101...17A} Arnaud, K.~A.\ XSPEC: The First Ten Years. Astronomical Data Analysis Software and Systems V, 101, 17 (1996)

\bibitem[40]{2012ApJ...756..128F} Foster, A.~R., Ji, L., Smith, R.~K., \& Brickhouse, N.~S.\ Updated Atomic Data and Calculations for X-Ray Spectroscopy. \apj, 756, 128 (2012)
  
\bibitem[41]{1989GeCoA..53..197A} Anders, E., \& Grevesse, N.\ Abundances of the elements - Meteoritic and solar. \gca, 53, 197 (1989)

\bibitem[42]{2003ApJ...597L.121E} Eisenhauer, F., Sch{\"o}del, R., Genzel, R., et al.\ A Geometric Determination of the Distance to the Galactic Center. \apjl, 597, L121 (2003)

\bibitem[43]{2000A&A...356..788C} Churazov, E., Forman, W., Jones, C., \& B{\"o}hringer, H.\ Asymmetric, arc minute scale structures around NGC 1275. \aap, 356, 788 (2000)

\bibitem[44]{2012ARA&A..50..455F} Fabian, A.~C.\ Observational Evidence of Active Galactic Nuclei Feedback. \araa, 50, 455 (2012)

\bibitem[45]{2014ARA&A..52..529Y} Yuan, F., \& Narayan, R.\ Hot Accretion Flows Around Black Holes. \araa, 52, 529 (2014)

\bibitem[46]{2018PASJ...70R...1K} Koyama, K.\ Diffuse X-ray sky in the Galactic center. \pasj, 70, R1 (2018)

\end{thebibliography}
\end{document}